\documentclass[twocolumn,tighten]{aastex62}

\hypersetup{colorlinks,%
 citecolor=blue,%
 linkcolor=blue}

\definecolor{fgreen}{rgb}{1,0.6,0.1}

\received{\today}
\revised{\today}
\accepted{\today}
\submitjournal{AJ}

\shorttitle{Dwarf-Dwarf merger}
\shortauthors{Paudel et al.}
\begin{document}
\title{KUG 0200-096: Dwarf antennae hosting a tidal dwarf galaxy}

\author[0000-0003-2922-6866]{Sanjaya Paudel}
\affil{Department of Astronomy and Center for Galaxy Evolution Research, Yonsei University, Seoul 03722, Korea}
\email{sanjpaudel@gmail.com (SP)\\
sengupta.chandreyee@gmail.com (CS)\\
sjyoon0691@yonsei.ac.kr (SJY)}
\author{Chandreyee. Sengupta}
\affiliation{Department of Astronomy, Yonsei University, 50 Yonsei-ro, Seodaemun-gu, Seoul, Korea}
\author{ Suk Jin Yoon}
\affiliation{Department of Astronomy and Center for Galaxy Evolution Research, Yonsei University, Seoul 03722, Korea}

\begin{abstract}

We study a gas rich merging dwarf system KUG 0200-096. Deep optical image reveals an optically faint tail with a length of 20 kpc, giving a visual impression of tidal antenna similar to NGC 4038/39. The interacting dwarf galaxies have B-band absolute magnitudes -18.06 and -16.63 mag. We identify a young stellar clump of stellar mass of 2$\times$10$^{7}$ M$_{\sun}$ at the tip of the antenna, possibly a Tidal Dwarf Galaxy (TDG). The putative TDG candidate is quite blue with $g-r$ color index of -0.07 mag, whereas the interacting dwarf galaxies have $g-r$ color indices 0.29 and 0.19 mag. The TDG is currently forming star at the rate of 0.02 M$_{\sun}$/yr.  We obtained HI 21 cm line data of KUG 0200-096 using the GMRT to get a more detailed view of neutral hydrogen (HI) emission in interacting dwarf galaxies and its TDG. Evidence of merger between the dwarf galaxy pair is also presence in HI kinematics and morphology where we find the HI contents of interacting pair is disturbed, forming a  extended tail toward the TDG. The HI velocity field shows strong gradient along the HI tidal tail extension. We present a comparative study between the  Antennae galaxy, NGC 4038/39, and KUG 0200-096 in both optical and HI gas properties and discuss possible origin of  KUG 0200-096 TDG.
\end{abstract}

\keywords{galaxies: dwarf,  galaxies: evolution galaxies: formation - galaxies: stellar population - galaxy cluster: Virgo cluster}

\section{Introduction}

The Antennae (NGC 4038/39) is the text book example of a pair of merging disk galaxies \citep{Arp66,Struck99}. As such, they have been studied in detailed observation  \citep{Amram92,Kunze96,Neff00,Gao01,Gordon01,Hibbard01,Whitmore05,Zhang10,Whitmore14} and reproduced in various numerical simulations \citep{Toomre72,Barnes88,Karl10,Teyssier10,Renaud15,Lahen18}. The system displays a prominent pair of tidal tails which extend a projected distance of 20' and the two merging disks are visibly distinct \citep{Schweizer78}. The latter has been assumed to be an indication of an early merger state, putting the system in the first place of the \cite{Toomre77} merger sequence. One important aspect of the Antennae system has been the discovery of ongoing formation of tidal dwarf galaxies at the tip of the antenna \citep{Mirabel92,Hibbard01}.

Tidal Dwarf Galaxies (TDGs) are the most massive sub-structures born in gas-rich mergers. Their total mass is typical to that of dwarf galaxies (M$_{*}$ $<$ 10$^{9}$). Made out of tidal material (gas and stars) ejected from galaxies into the intergalactic medium, they are independent gravitationally bound systems, usually supported by rotation. Their dynamical status qualifies them as ``galaxies" \citep[see][for a review on TDGs]{Duc12} although being nascent galaxies they are not necessarily always in dynamical equilibrium \citep{Lelli15}.  The importance of TDGs among the dwarf population is rather controversial. The idealized numerical simulations of galaxy-galaxy collisions made by \cite{Bournaud06} suggests that only a fraction of massive TDGs might survive longer enough to evolve as independent galaxies. Objects that are not kicked out from their parent's potential well are subjected to dynamical friction and gradual reduction of orbital energy plunges them into the host system where they could suffer destructive tidal force of the host galaxies \citep{Mayer01,Fleck03}, or can be destabilized by effect of ram pressure \citep{Smith13}.

While massive galaxy interactions have been studied in great detail in the past, very little is known about the evolution of dwarf-dwarf interactions and mergers. There has been growing interest in the dwarf-dwarf interaction in recent literatures and in the last few years number of studies have presented observational evidences of the merging dwarf galaxies \citep{Paudel15, Stierwalt15,Pearson16,Delgado12}. In addition, many star-bursting dwarf galaxies show disturbed HI kinematics as a signature of tidal interactions \citep{Lelli14}. Studying formation TDGs in the dwarf-dwarf merging systems is important as the TDGs born out of colliding dwarf galaxies is expected to have different environment. Owing to the shallow potential well of host galaxies, the new born TDGs are subjected to a significantly low level of harsh tidal force from the parent. Low mass galaxies  also have lower level of X-ray emission therefore less strong ram pressure stripping effect. This provides a higher survival probability of TDGs born out of dwarf galaxy collisions.

\begin{table*}
\caption{Photometric properties}
\begin{tabular}{ccCc R cCcCccc}
\hline
Galaxy & RA & Dec & m$_{r}$  & g-r & m$_{FUV}$ & M$_{B}$ &  M$_{*}$& SFR &  M$_{HI}$ & z  & V$_{r}$\\
 & h:m:s & d:m:s & mag & mag & mag & mag & log(M$_{\sun}$) & log(M$_{\sun}$/yr)&  log(M$_{\sun}$) & &km/s  \\
\hline
D1   & 02:02:38.77 & -09:22:13.2 &  15.76 &  0.29 & 18.06 & -18.06  & 9.27 & -0.68 & 9.3 & 0.018 & 5376 \\
D2   & 02:02:39.95 & -09:22:43.0 &  17.32 &  0.19 & 19.28  & -16.63  & 8.74 & -1.17 & \nodata &\nodata & 5355 \\
TDG  & 02:02:39.14 & -09:23:23.4 &  20.36 & -0.07 & 20.62  & -13.93  &  7.27 & -1.70 & 8.1 & \nodata &  5441 \\
\hline
\end{tabular}
\label{phtab}
\\
The magnitudes are corrected for galactic extinction and the B-band magnitude is obtained from the SDSS $g-$band magnitude using conversion formula provided by the SDSS. Typical error on the SDSS magnitudes are 0.01 mag. The stellar mass, given in column 8, is derived from the $r-$band luminosity and the mass-to-light ratio obtained from \cite{Bell03} for the color $g-r$ where we expect an conservative error on our estimate is 0.2 dex. We list star-formation rate in column 9, which is estimated from the FUV magnitudes. In column 10, we give HI mass, were the D1 value represents HI mass of entire the system. Radial velocity measured from HI kinematics is listed in the last column.
\end{table*}

In this paper we present an unique example of a low mass (M$_{*}$ $\approx$4$\times$10$^{9}$) Antennae system, KUG 0200-096, where we identify a TDG at the tip of tidal tail. 

Throughout the paper we adopt a luminosity distance of 68 Mpc (m - M = 34.1) and a scale of 0.32 kpc/arcsec valid for H0 = 71 km/s Mpc$^{-1}$.
\section{KUG 0200-096: The Dwarf Antennae}
At a sky position Ra = 02:02:38.77 Dec = -09:22:13.2 and redshift z = 0.018, we find an interacting pair of star-forming dwarf galaxies, KUG 0200-096, with an extended tail of stellar stream.  According to NED query, KUG 0200-096 is located in an isolated environment. It's immediate neighbor galaxy is NGC 0787 at a sky-projected distance of 693 kpc and the two have relative line of sight radial velocity 692 km/s.  As far as our NED query results, we find no other companion, not even dwarf, galaxies around KUG 0200-096 within the area of 700 kpc and $\pm$1000 km/s velocity range. 
\begin{figure*}
\centering
\includegraphics[width=18cm]{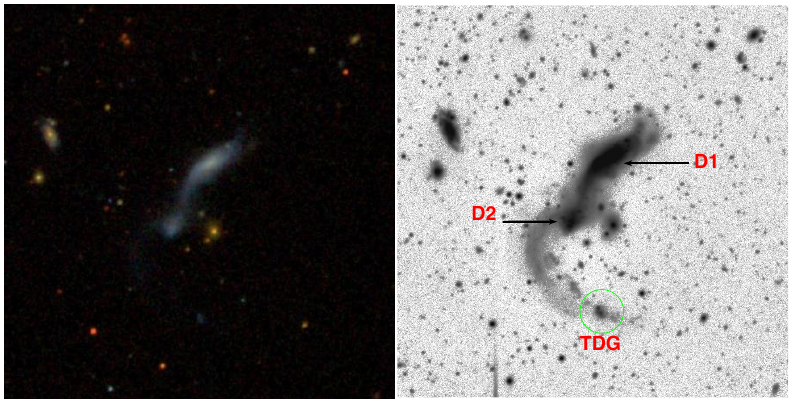}
\caption{Optical view KUG 0200-096 seen from the SDSS g-r-i combined tri-color image. In right, we show a gray scale image produced from much deeper CHFT archival g-band image. Both of them have similar field of view of 3\arcmin$\times$3\arcmin. We identify the putative TDG by a circle and mark D1 and D2 by arrows.}
\label{main}
\end{figure*}

Figure \ref{main} reveals the remarkable tidal nature at the periphery of KUG 0200-096. The color image is obtained from the SDSS sky-server which is prepared by combining $g$, $r$ and $i$-band images \citep{Lupton04}. In right, we show a much deeper gray scale $g$-band image which we have obtained from Canada France Hawaii Telescope (CHFT\footnote{http://www.cfht.hawaii.edu}) archive. Both have 3$\arcmin$$\times$3$\arcmin$ field of view. We label interacting pair of dwarf galaxies D1 and D2. The sky-projected separation between the two galaxies (D1 and D2) is 9 kpc. It is apparent that D2 is more disturbed than D1 forming an antenna like prominent tidal tail which mimics one antenna of  NGC 4038/39 (The Antennae galaxy). The shape of the antenna is nearly semi circle and surface brightness is not uniform. It decreases toward the end. In fact, this is almost invisible in the color SDSS image and we can only spot a blue clump, quite separated from the interacting dwarf pair main body.

We consider the blue clump at the tip of the antenna, highlighted by a green circle, as a potential TDG candidate. The left color image reveals that the TDG is much bluer compared to the D1 and D2 which, indeed, will be verified by measured $g-r$ color, see below. TDG position in they sky is Ra = 02:02:39.14 and Dec = -09:23:23.4. It is well separated from the interacting galaxies main bodies with distances from D1 and D2 of 22 and 14 kpc, respectively.

\section{Data analysis}\label{data}
\subsection{Analysis of optical data}
To perform a photometric analysis and measure the total luminosity, we retrieved archival images from the SDSS-III database \citep{Abazajian09}. Since the SDSS-III data archive provides well calibrated and sky-background subtracted image, no further effort has been made in this regard. However, we used a simple and similar approach to subtract the sky-background count as in \cite{Paudel14b}.   We performed the aperture photometry with different aperture size to measure total flux of objects of interest, i.e  D1, D2 and the potential TDG candidate. The sizes of apertures were selected visually where we used a wide enough aperture that secure all the flux in the region of interest. Before the doing aperture photometry, we masked all unrelated foreground and background objects manually.

We list the positions, absolute magnitudes, and stellar masses of the interacting dwarf member and the putative TDG in Table \ref{phtab}. The derived magnitudes were corrected for Galactic extinction using \cite{Schlafly11}, but not for the internal extinction. We convert  the SDSS  $g$-band magnitudes to B-band magnitude using equation provided by the Lupton (2005), available in the SDSS website\footnote{http://www.sdss3.org/dr8/algorithms/sdssUBVRITransform.php}.

To estimate the stellar masses, we use a calibration provided by \cite{Bell03} appropriate for the observed $g-r$ color. \cite{Bell03} assume a simple stellar population with a single burst of star-formation to derive a relation between mass-to-light ratio and galaxy color where a typical scatter is 0.2 dex. Indeed, galaxy star-formation histories are complex.  However, it is shown that scatter in the mass-to-light ratio derived from different star-formation history for a given color is also scattered within $\pm$ 0.2 dex, see \S4.3 \cite{Zhang17}. Therefore, we consider 0.2 dex as a typical conservative error on our stellar mass estimates.

The interacting dwarfs, D1 and D2, have B-band absolute magnitude -18.06 and -16.63 mag, respectively. We find that both interacting member dwarf galaxies have approximately similar $g-r$ color index, i.e 0.29 and 0.19 for D1 and D2, respectively. But the TDG is significantly blue, with $g-r$ color index -0.07 mag. The total stellar mass of interacting dwarfs D1 and D2 and the TDG candidate are 1.9$\times$10$^{9}$, 5.5$\times$10$^{8}$ and 1.9$\times$10$^{7}$ M$_{\sun}$, respectively. That gives a stellar mass ratio between the interacting dwarfs is 3:1. 

\begin{figure}[h]
\includegraphics[width=8cm]{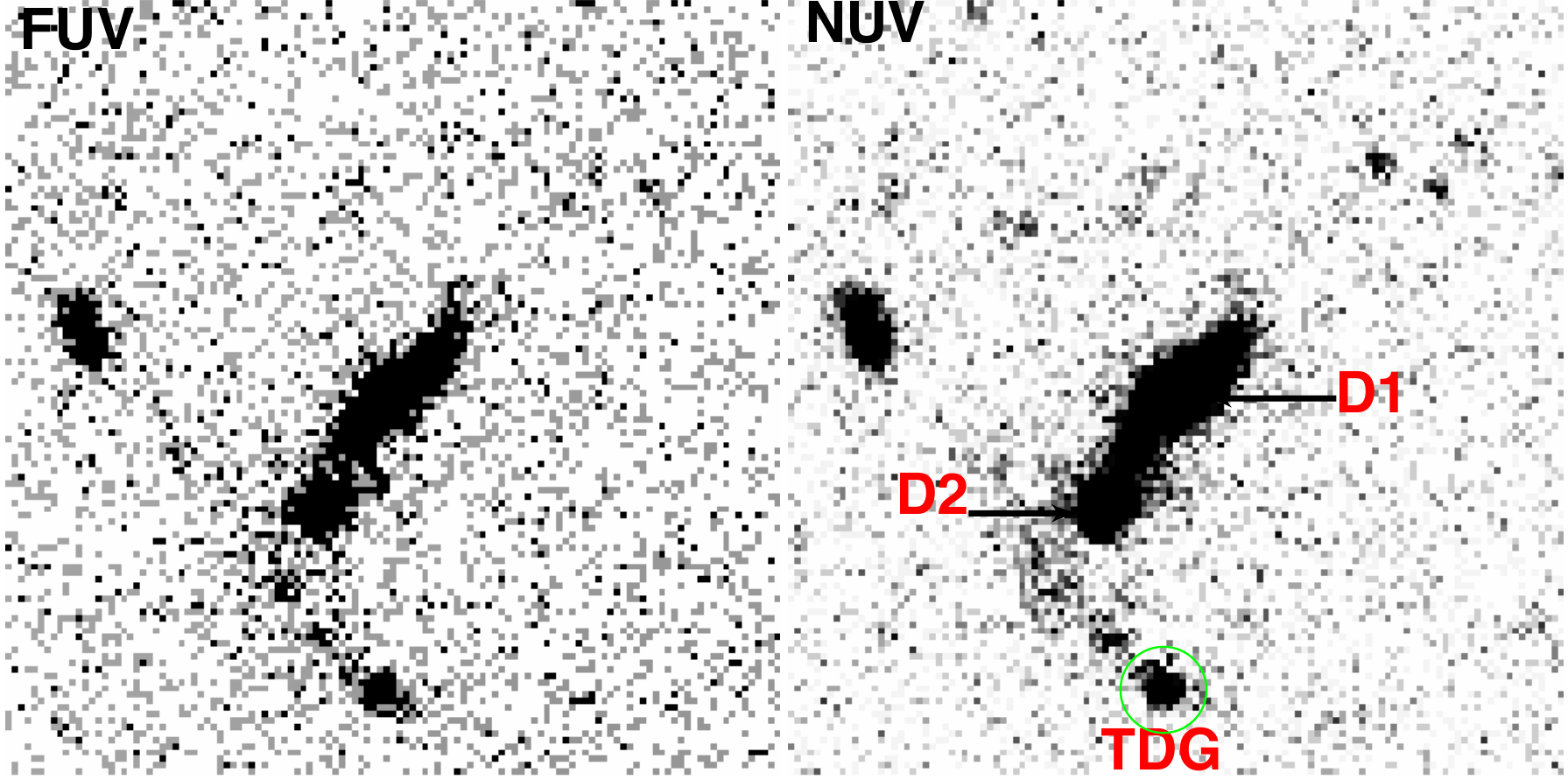}
\caption{GALEX all-sky survey FUV (left) and NUV (right) images of KUG 0200-096.}
\label{uvpic}
\end{figure}

As their blue color indicates ongoing star-forming activity, we use  GALEX UV images to calculate ongoing star-formation rates. We retrieved The GALEX all-sky survey \citep{Martin05} archival images, see Figure \ref{uvpic}. Interestingly, the stellar stream/antenna is also visible in GALEX UV images and the TDG is quite prominent in both FUV and NUV bands. We perform aperture photometry as done in the SDSS images in both FUV and NUV-band images. We derive the star-formation rate from the FUV flux using a calibration provided by \cite{Kennicutt98}. The values are listed in Table \ref{phtab}. UV-optical color (FUV$-r$) of D1 and D2 are 2.3 and 1.9 mag, respectively, and that of the TDG is significantly blue, i.e 0.26 mag.

Finally, we performed aperture photometry on the entire system. For this, we first manually masked all non-related objects and selected a large aperture that includes the entire system. The total B-band luminosity is  M$_{B}$ =  -18.69  and the $g - r$ color index is 0.29 mag. We derived a total stellar mass M$_{*}$,total = 3.9$\times$10$^{9}$M$_{\sun}$ and a total star formation rate is SFR$_{total}$ = 0.46 M$_{\sun}$ /yr. Form this estimates, we calculate the TDG mass fraction is only $\approx$0.5\%  of entire system. A significant increase in the stellar mass of the entire system in comparison with sum of the stellar masses of three components, D1, D2 and TDG, could be because of following three reasons. 1) we may have missed some light from extended parts, particularly stellar stream which are only accounted during overall photometry using a single large aperture. 2) there might be contamination from foreground and background structure which we have misidentified and 3) different star-formation history which produces slightly different mass-to-light ratio. However, we expect that all three might have contributed to some extent to increase the stellar mass of the overall system.

The SDSS fibre spectroscopy has targeted the brighter galaxy, D1, and gives its redshift z = 0.01801. We obtain the optical spectrum of D1 from the the SDSS archives which is observed with a 3\arcsec diameter fibre. It exhibits the emission lines typical of HII regions as well as relatively strong absorption for the early Balmer lines H$\delta$, $\gamma$ and $\beta$. H$\alpha$ equivalent width measured in the SDSS fibre spectrum is 22 \AA{} and this value is not high enough to consider the galaxy as a star-burst, this is typical for local Blue Compact Dwarf galaxies \citep[BCDs][]{Meyer14}. Emission line metallicity derived from ration of H$\alpha$/[NII] is 8.3 and star-formation rate derived from H$\alpha$ emission line flux is 0.015 M$_{\sun}$/yr which is significantly lower than SFR derived from FUV flux. However note that  3\arcsec diameter fibre spectroscopy only represent the central part of the galaxy and SFR derived from FUV flux is summed over entire galaxy.

\subsection{Radio 21-cm observation}\label{radio}
\begin{figure}
\includegraphics[width=8cm]{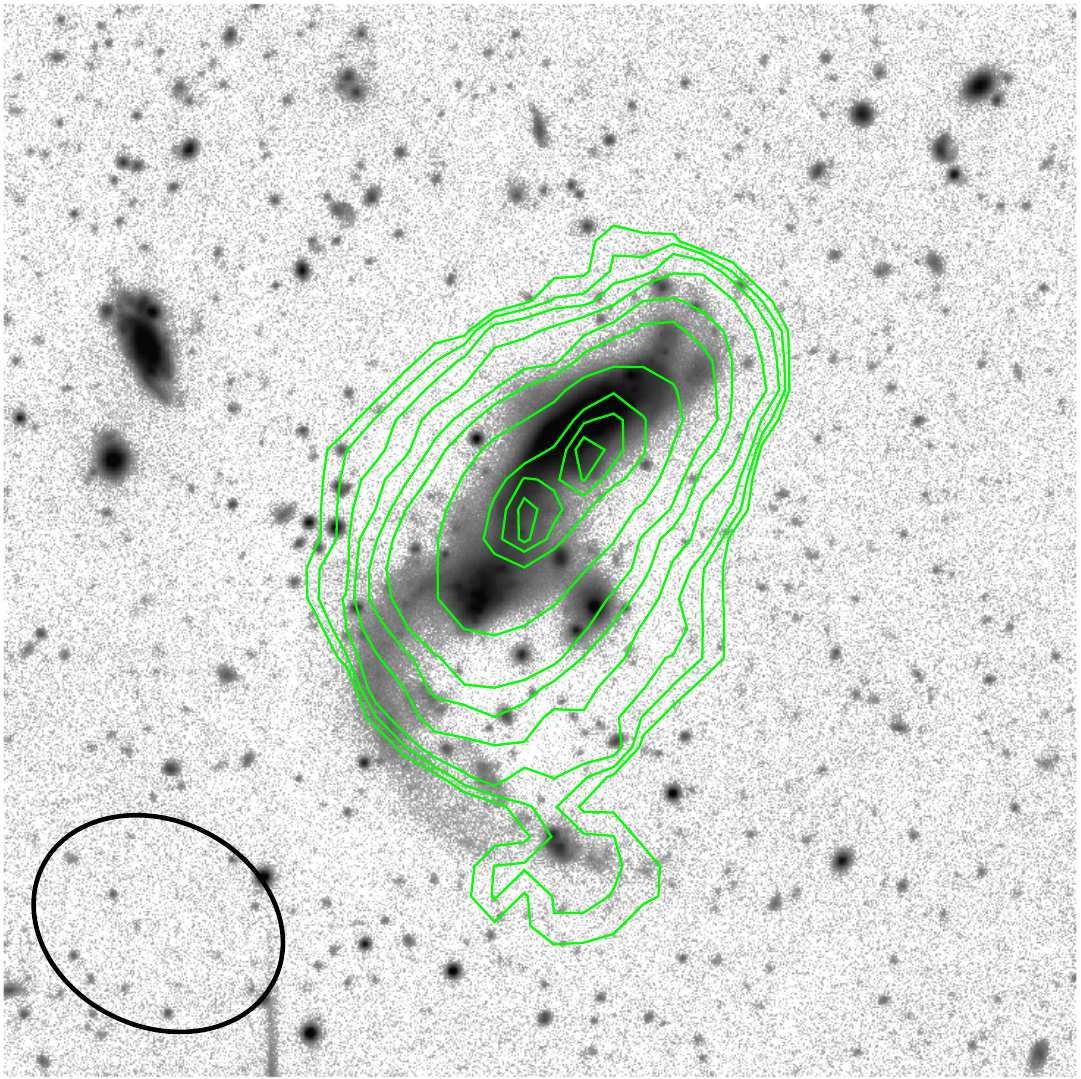}
\includegraphics[width=8cm]{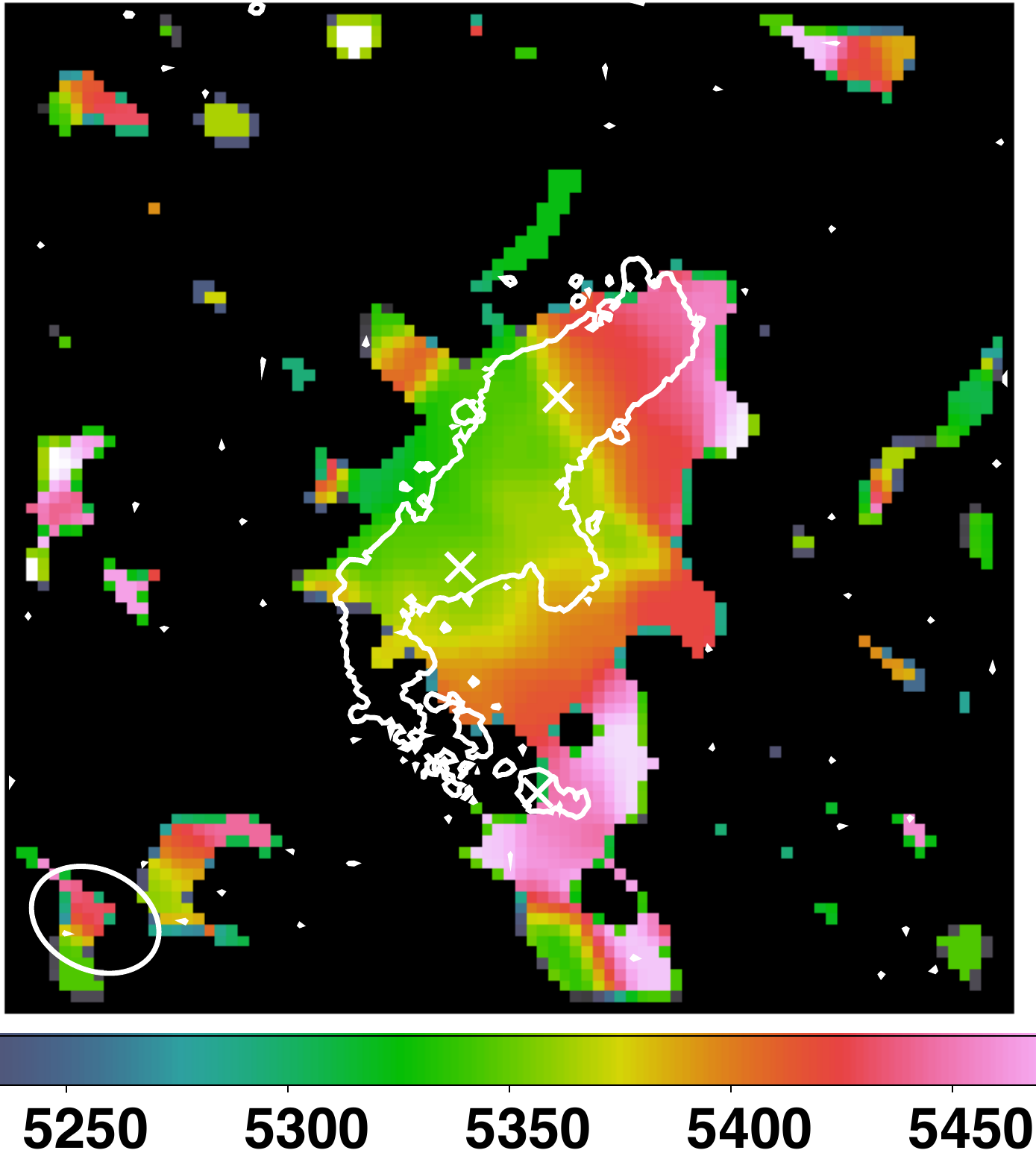}
\caption{Top: Integrated H I contours from the GMRT low-resolution map overlaid on the CHFT  g-band image. The H I column density levels are N(H I) = 10$^{19}$ (2.4, 4.1, 6.6, 9.9, 13.3, 16.6, 19.9, 23.3, 26.7, 30.0) cm$^{-2}$. A black ellipse at the bottom left corner represents beam size of GMRT observation.
\newline Bottom: we show the H I velocity field from the GMRT high-resolution cube for H I emission $>$3$\sigma$. The white contour traces the optical g-band light at the level  surface brightness 26 mag arsec$^{-2}$. The white crosses represent the position of D1, D2 and TDG.}
\label{himap}
\end{figure}

\begin{figure*}
\centering
\includegraphics[width=17cm]{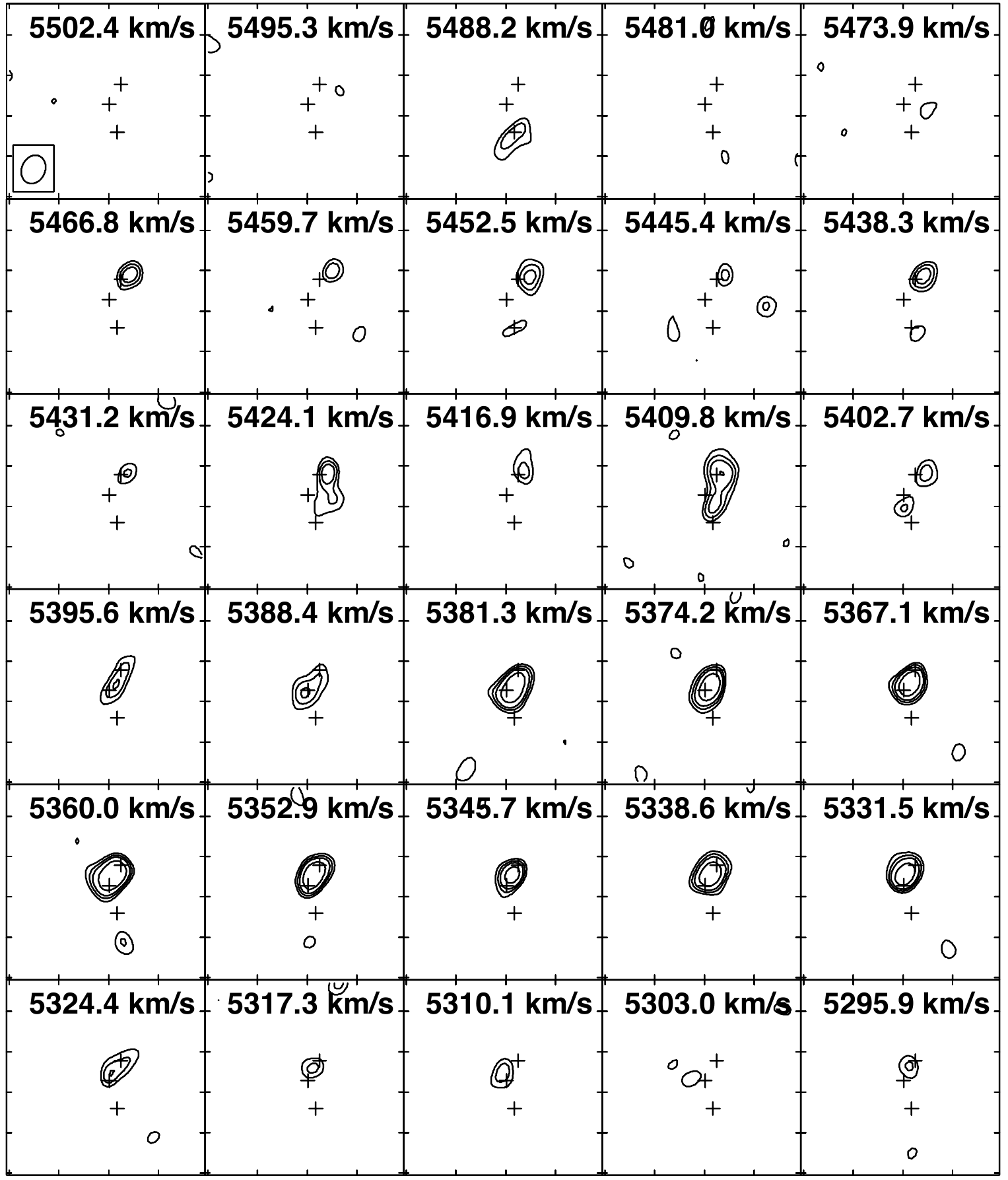}
\caption{The HI velocity-channel contour map of KUG 0200-096. The contours are shown at  1.1$\times$10$^{-03}$$\times$(3, 4, 5, 7, 9) Jy/beam. The northern, eastern and southern cross marks represent of the position of D1, D2 and TDG, respectively. Observed beam size is shown in top left panel.}
\label{chmap}
\end{figure*}

To study the HI content of the system in detail, we carried out  HI interferometric observations of KUG 0200-096, using the Giant Metrewave Radio Telescope (GMRT\footnote{http://www.ncra.tifr.res.in/ncra/gmrt}) located at Pune, India.  The system was observed on June 15th, 2017 as part of our observing proposal ``Formation of Tidal Dwarf Galaxies (TDG) during dwarf-dwarf merger". Part of the data from this project has already been published in \citep{Paudel17c} where we the describe observation setup. In brief, a baseband bandwidth of 16 MHz was used for the observations yielding a velocity resolution $\sim$7 km/s. The GMRT primary beam  at L band is 24$^{\prime}$ and the synthesised beams  of the images presented in the paper are 43.4$^{\prime\prime}$ $\times$ 34.6$^{\prime\prime}$ (low resolution) and 24.8$^{\prime\prime}$ $\times$ 19.5$^{\prime\prime}$ (high resolution).  At the adopted distance of KUG 0200-096, 68 Mpc, 43\arcsec and 25\arcsec sample 14 kpc and 8 kpc respectively. The data was analysed using the software {\tt AIPS}\footnote{http://www.aips.nrao.edu} and the procedure followed was similar to that explained in \cite{sengupta17}.

Figure \ref{himap}, top panel shows the contours from the low resolution integrated HI map overlaid on the CHFT $r$-band image.  The HI emission is mainly concentrated around D1 and D2 with an extension of tenuous emission in the direction of antenna.  An HI tail extends towards the antenna approximately aligned to the stellar stream. Except the HI tail, we do not detect any extended or diffuse HI features around the system. The HI column density at the tail  region is very low, $\sim$2-4 $\times$10$^{19}$ cm$^{-2}$.

The HI map reveals two peaks in HI column density suggesting both the interacting galaxies may be gas rich. The misalignment of the optical and HI tidal remnants is more pronounced in the high resolution map compared to the low resolution map, see lower panel Figure \ref{himap}. The HI is  seen to overlap with the TDG forming a bridge to the main system with no optical counterpart. On the TDG,  the peak HI column density is 4$\times$10$^{19}$ cm$^{-2}$. The TDG is quite compact in optical, but our resolution limit of HI observation does not allow us to conclude whether the gas structure in TDG is already detached or still part of the  tidal tail.

\begin{figure}
\includegraphics[width=8cm]{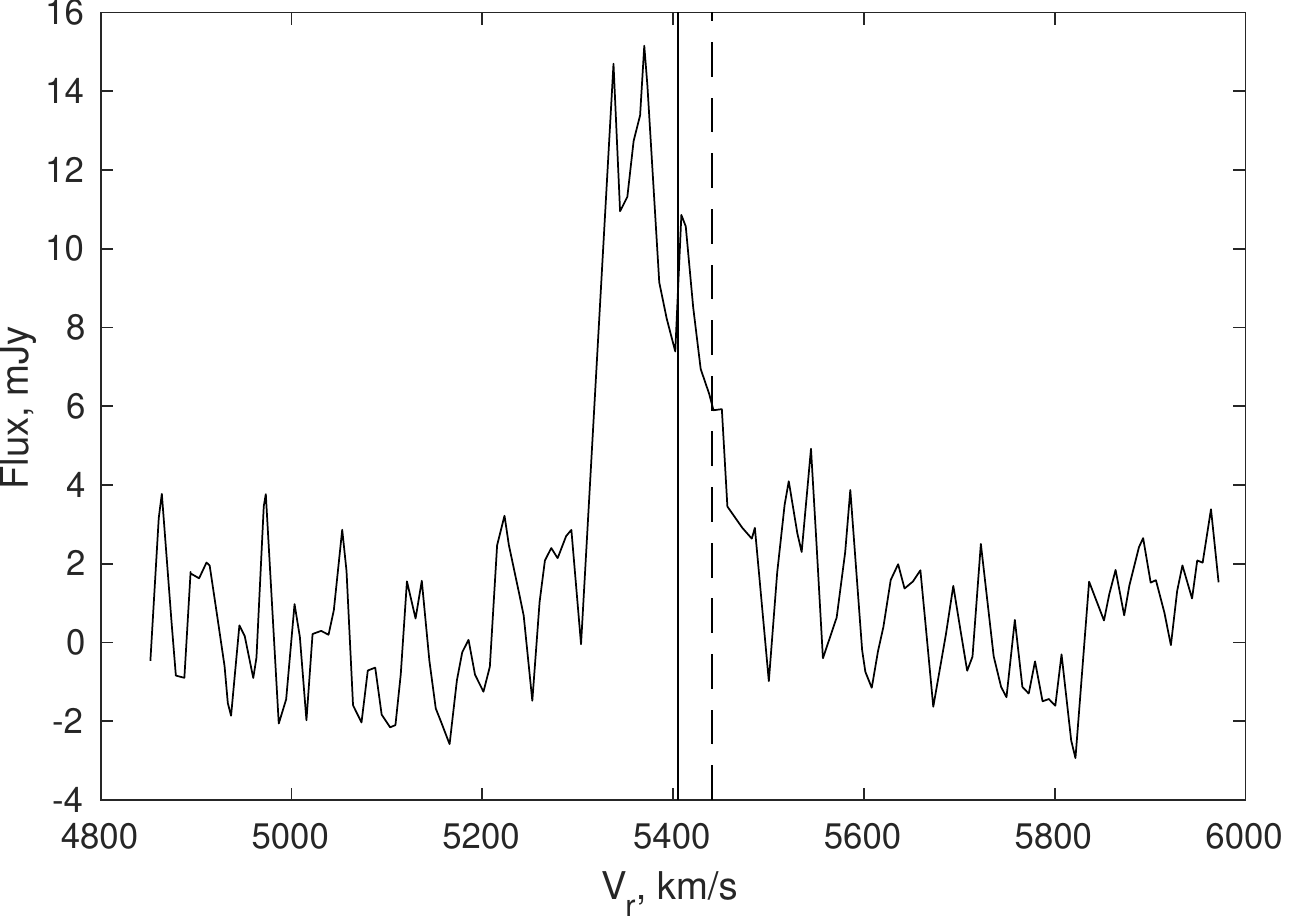}
\caption{Integrated HI spectrum. We draw two vertical lines, slid and dash, to show optical velocity of D1 and the HI velocity of the TDG, respectively. }
\label{tspec}
\end{figure}

Figure  \ref{himap} bottom panel shows the HI velocity field of the KUG 0200-096 system in high resolution. While signs of an interacting system are clear, velocity fields of the individual galaxies (D1 and D2) cannot be resolved due to lack of spatial resolution.  However, we can infer that D2 has lower line of sight velocity than D1. Figure \ref{tspec} shows the integrated spectrum of the combined system where we mark the optical velocity of D1 in solid line and the HI velocity of the TDG in dashed line. The TDG HI velocity represents the peak HI velocity in the HI spectrum of region around TDG. Interestingly, the gas around the TDG has velocities closer to D1.

The HI tidal tail structure and kinematics are further illustrated in the channel maps, shown in Figure \ref{chmap}. The northern, eastern and the southern crosses represents D1, D2 and the TDG, respectively. HI emission is detected in the system, in the velocity range of 5296 to 5488 km/s. The origin of the HI connection of the TDG to the parent interacting pair becomes clearer in the channel maps. While the optical image shows the stellar tidal tail to originate in D2, the HI channel maps indicate the HI tidal tail connecting the TDG to KUG 0200-096 originates at D1, between velocities 5402 to 5488 km/s. The integrated HI flux density of the system is  1.4 Jy km/s and in the TDG region is 0.4 Jy km/s which corresponds to  HI mass of 1.5 $\times$ 10$^{9}$ and 4 $\times$ 10$^{8}$  M$_{\sun}$, respectively.  HI gas-to-light ratio, log(M$_{HI}$/L$_{B}$) for the TDG and overall system is 0.4 and -0.3, respectively.

\section{Discussion}
We have presented the case of a merging dwarf galaxies pair in an isolated environment where we found a TDG is forming at the tip of tidal stream similar to the well known interacting system NGC 4038/39 (The Antennae galaxy). The TDG has a stellar mass M$_{*}$ = 1.9$\times$10$^{7}$ M$_{\sun}$ which is 0.5 percent of entire merging system. It is located at 22 kpc sky projected distance from main merging galaxy. It is blue with $g-r$ color index of -0.07 mag and gas rich with HI gas-to-light ratio of log(M$_{HI}$/L$_{B}$) = 0.4.

\subsection{Tidal interaction and Star-formation}
KUG 0200-096, no doubt, provides a great example of gas-rich merging dwarf galaxies. The interacting galaxies, D1 and D2 have B-band absolute magnitudes -18.06 and -16.63 mag which is similar to that of LMC/SMC pair, a well known interacting dwarf galaxies in our near vicinity. Several physical properties of KUG 0200-096, i.e., color, metal content, and SFR, are fairly similar to the typical BCDs and there is little doubt that its star formation activity is affected, if not triggered, by the interaction.

LMC/SMC do not host star-forming region out-side of galactic main body, although it shows substantial extension of gas structure  \citep{Onghia16} . However, in our previous publication \citep{Paudel17}, we have identified a TDG located in a stellar bridge between two interacting dwarf galaxies, which have similar star-formation and physical properties to LMC/SMC.

\begin{figure}
\includegraphics[width=8cm]{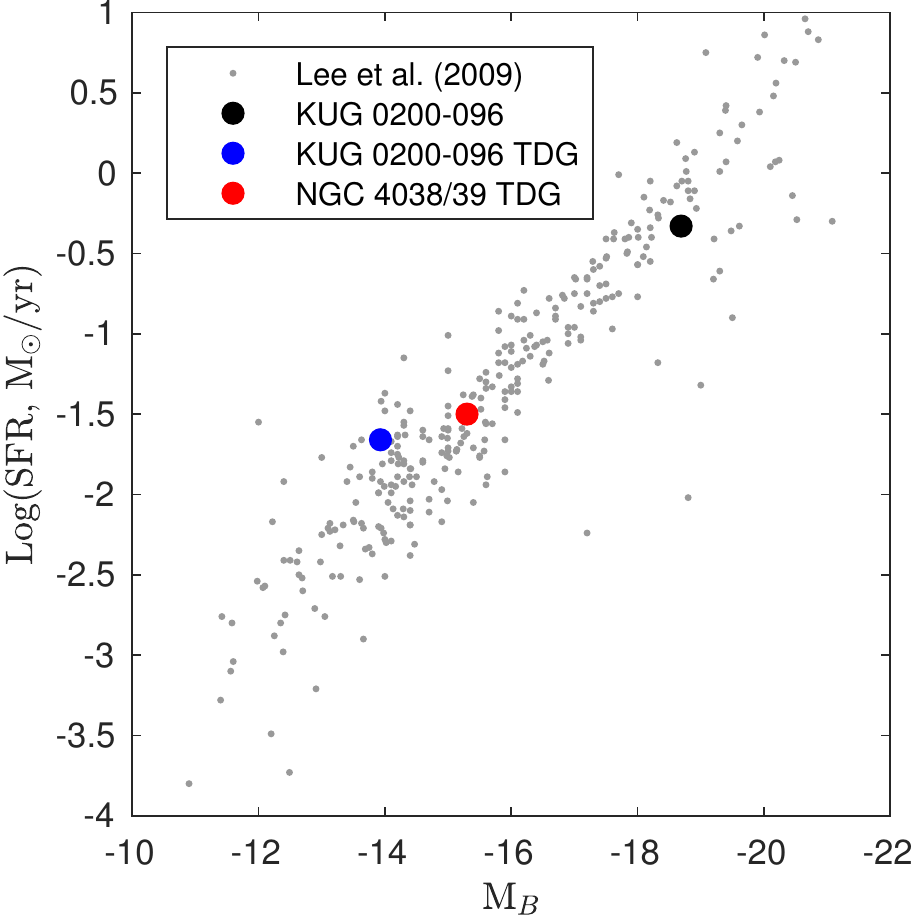}
\caption{Relation between SFR and  B-band absolute magnitude. The black dot represents overall KUG 0200-096 and blue dot for its TDG. NGC 4038/39 TDG is shown in red dot. The comparison sample, gray dots, is taken from \cite{Lee09}. }
\label{sfr}
\end{figure}

We find that on average KUG 0200-096 has typical star-forming properties of BCDs. In Figure \ref{sfr}, we show a relation between star-formation rate and B-band magnitude for star-forming galaxies. The comparison sample is taken from \cite{Lee09},  who study star forming activity  of local volume ($<$11 Mpc) star-forming galaxies where star-formation rates are also derived from FUV flux. We find no enhanced star-formation for overall B-band magnitude of the system compared to a sample of star-forming galaxies from the local volume. However, the TDG is located at the upper edge of scattered data. 

The gas mass fraction of the TDG is high compared to the overall gas mass fraction of the system but due to large beam size of GMRT observation, we can not rule out that the possibility of some contamination from the host galaxies. In any case, the value of gas to stellar mass ratio of the TDG, log(M$_{HI}$/M$_{*}$) = 0.83, is perfectly scaled with the relation of log(M$_{HI}$/M$_{*}$) and stellar mass, see \cite{Popping15} Figure 3.

The elongated tails are a clear sign of tidal interaction of nearly equal-mass gas rich-disk galaxies. According to Toomre sequence \citep{Toomre72}, KUG 0200-096 is probably in early-stage of interaction. In comparison to the Antennae system, both interacting galaxies are clearly separated which may hint that the interaction in KUG 0200-096 is young compare to the interaction between NGC 4038 and NGC 4039. 

Numerical simulations of this interacting system could help to assess time scale of interaction and further morphological evolution of the TDG and merging remanent but they are beyond the scope of this paper.

\subsection{Formation of TDGs during dwarf-dwarf merger}
Merging probability of low mass galaxies decreases in low redshift universe and as a result  the chance of formation of tidal dwarf galaxies by merger of dwarf galaxies is also low \citep{Lucia06}. This makes low redshift dwarf-dwarf mergers an interesting phenomena to study. It has been found that dwarf-dwarf interactions are more likely to happen in isolated environments than in the groups or clusters  \citep{Stierwalt15,Paudel18}. Given that dwarf galaxies located in isolated environment are gas rich, mostly Blue Compact Dwarf galaxy (BCD) type, it is not  surprising that TDGs may form  frequently in these gas rich dwarf-dwarf mergers. In fact, we also identified a new born TDG in our previous study of merging dwarf galaxies although they are in group environment \citep{Paudel15,Paudel17c}.

KUG 0200-096 is noteworthy for the presence of a stellar stream hosting clump of star formation at the tip, and in that respect it resembles the system involving massive colliding galaxies, such as the Antennae  (NGC 4038/39). In the study of numerous massive interacting galaxies, evidence of in situ star formation occurring in gas-rich collisional debris has been reported \citep{Mundell04,Mello08,Peterson09}. Such regions are also believed to be a nursery of super star clusters or TDGs \citep{Duc94,Duc07,Duc14,Paudel15,Paudel17}. These evidences of observation are also supported by idealized and cosmological  numerical simulations of galaxies, where massive and compact super star clusters are seen forming in tidal tails \citep{Bournaud08,Renaud15,Ploeckinger18} and some of them may have evolved independently and survive against internal feedback and external tidal shear. The most massive and extended of them may become independent TDGs \citep{Wetzstein07}.

The same phenomena seem to also occur in dwarf-dwarf major mergers.  We find  a blue stellar clump at the tip of stellar stream which hosts star forming region. It has a stellar mass of 1.9$\times$10$^{7}$ M$_{\sun}$. As its parent galaxies are low-mass systems with shallow potential wells, one may speculate that it will survive longer than in an environment of systems involving massive merging galaxies, e.g. NGC 4038/39.

In comparison, NGC 4038/39 TDG is brighter than KUG 0200-096 TDG with a V-band absolute magnitude of -15.3 mag and has star-formation rate 0.03 M$_{\sun}$/yr \citep{Mirabel92}. Currently,  KUG 0200-096 TDG is forming stars at the rate of 0.02 M$_{\sun}$/yr and both follow a the scaling relation of SFR and blue band absolute magnitude defined by normal galaxies, see Figure \ref{sfr}.  \cite{Hibbard01} presented a high resolution HI mapping of NGC 4038/39 and its TDG Candidates. They found the HI morphology possess plenty of tidal features and substructures. The HI tail nearly follow the extension of both antennae observed in the optical. In KUG 0200-096, we find that gaseous tail does not overlap with the optical counterpart. However note that our spatial resolution is not sufficient enough to resolve the HI tail, to confirm whether it has a substructure.

The TDG HI velocity is closer to that of  D1 and a careful examination of Figure \ref{chmap} reveals that the extended HI emission towards the TDG actually originates in D1 (see higher velocity channel maps). It is possible that the HI tail actually emerges from D1, and that the stellar stream and the HI extension do not have the same point of origin, but are just projected on each other on sky. That can also explain the observed offset between the HI tail and the stellar stream and the lower relative line of sight velocity between D1 and the TDG. In this scenario, the TDG may be located at the end of the HI tail but the location of the TDG near the tip of stellar stream maybe a chance projection.  Like the Antennae, KUG 0200-096, may be hosting two antennae -one is the gas poor stellar stream originated from D2 and another is HI tail originated from D1. In any case, to confirm this we certainly need to study a higher resolution and better signal to noise ratio HI map.

Probably, most interesting difference between NGC 4038/39 TDG and KUG 0200-096 TDG is that the latter is significantly more compact compared to the former. NGC 4038/39 system is more massive and the TDG has a 15 kpc diameter whereas KUG 0200-096 TDG diameter is 2.5 kpc. \cite{Weilbacher18} identify multiple sub-clumps in the NGC 4038/39 TDG and detected multiple HII regions. In that sense, KUG 0200-096 TDG is morphologically more similar to BCDs, typically xBCDs \citep{Drinkwater91} and  in contrast morphological properties of NGC 4038/39 TDG is comparable to a typical dwarf irregular galaxy (dIr).

\acknowledgements
P.S. acknowledges the support by Samsung Science \& Technology Foundation under Project Number SSTF-BA1501-0. S.-J.Y. acknowledges support from the Center for Galaxy Evolution Research (No. 2010-0027910) through the NRF of Korea and from the Yonsei University Observatory -- KASI Joint Research Program (2018). 

This study is based on the archival images and spectra from the Sloan Digital Sky Survey (the full acknowledgment can be found at http://www.sdss.org/ collaboration/credits.html). We also made use of archival data from Canada-France-Hawaii Telescope (CFHT) which is operated by the National Research Council (NRC) of Canada, the Institute National des Sciences de l'Univers of the Centre National de la Recherche Scientifique of France and the University of Hawaii. We thank the staff of the GMRT that made these observations possible. GMRT is run by the National Centre for Radio Astrophysics of the Tata Institute of Fundamental Research.

\facility{GMRT} 

%\section{Conclusion}

%\bibliographystyle{aasjournal}
%\bibliography{ddref}

\end{document}